\documentclass[fleqn,preprint,showpacs,preprintnumbers,amsmath,amssymb]{revtex4-1}

\usepackage{graphicx}% Include figure files
\usepackage{dcolumn}% Align table columns on decimal point
\usepackage{bm}% bold math
\usepackage{xcolor}
\def\apj{{\em The Astrophsical Journal}}

\def\apjl{{\em The Astrophsical Journal Letters}}

\def\jgr{{\em J. Geophys. Res.}}
\def\prl{{\em Phys. Rev. Lett.}}

\def\pop{{\em Phys. Plasma}}

\def\solphys{{\em Solar Physics}}

\begin{document}
\preprint{APS/123-QED}
\title{Bi-Directional Energy Cascades and the Origin of Kinetic Alfv\'enic and Whistler Turbulence in the Solar Wind}% Force line breaks with \\

\author{H. Che}
\author{ M. L. Goldstein }
\author{A. F. Vi\~nas}
\affiliation{ NASA/Goddard Space Flight Center, Greenbelt, MD, 20771, USA}%Lines break automatically or can be forced with \\

%\date{\today}% It is always \today, today

\begin{abstract}
The observed ion-kinetic scale turbulence spectrum in the solar wind raises the question of how that turbulence originates. Observations of keV energetic electrons during solar quiet-time suggest them as possible source of free energy to drive kinetic turbulence. Using particle-in-cell simulations, we explore how the free energy released by an electron two-stream instability drives Weibel-like electromagnetic waves that excite wave-wave interactions. Consequently, both kinetic Alfv\'enic and whistler turbulence are excited that evolve through inverse and forward magnetic energy cascades.
\end{abstract}

\maketitle
The observations of solar wind turbulence have shown that at scales approaching the ion kinetic scale where the ions and electrons are decoupled and the kinetic effects must be considered, the power-spectrum of  magnetic fluctuations, which in the inertial range follows the Kolmogorov scaling $\propto k^{-5/3}$, is replaced by a steeper \cite{Leamon_et_al_1998a,Leamon_et_al_1999,Leamon_et_al_2000} anisotropic scaling law $B^2_{k_\perp} \propto k_{\perp}^{-\alpha}$, where $\alpha$ is a number larger than $5/3$. It is found that the observed spectral index is $\alpha \approx 7/3$, but this value is not universal and  varies from interval to interval.  Magnetic fluctuations with about tenth of ion gyro-frequency propagating nearly perpendicularly to the solar wind magnetic field are identified as kinetic Alfv\'enic waves (KAWs) \cite{bale05prl,sah09prl,kiyani09prl,alex09prl,salem12apjl,pod13solphys} and the break frequencies of the magnetic power-spectra appear to follow the ion inertial length \cite{Leamon_et_al_2000,perri10apjl,bou12apj}. The origin of the KAW turbulence is still unknown.  In this letter, we address the origin of kinetic turbulence by proposing a source of free energy that has not been explored previously. For the first time we find that an inverse energy cascade appears to play a crucial role in generating both KAW and Whistler turbulence.

Observations using the STEREO spacecraft have found that even during quiet-time periods, the solar wind contains a previously unknown electron population different from the core solar wind, called  ``superhalo electron", with energy ranging in $\sim 2- 20$ keV \cite{lin97conf,wang12apjl}. One possible origin of the superhalo electrons is the escaping nonthermal electrons related to coronal nanoflares in the quiet solar atmosphere (Parker 1988 \cite{parker88apj}; Lin 1997 \cite{lin97conf}). The relative drift of these nonthermal electrons to the background solar wind can drive an electron two-stream instability in a neutral current \cite{gary93book}, and release the free energy to the solar wind. The impact of this unstable process on the solar wind has so far not been studied. In this letter, using particle-in-cell (PIC) simulations, we investigate how the rapidly released energy drives Weibel-like electromagnetic waves. The wave-wave interactions on ion inertial scales $d_i=c/\omega_{pi}$ and electron inertial scales $d_e=c/\omega_{pe}$ generate KAW and whistler turbulence through both forward and inverse energy cascades. At the end of this letter, we will compare the testable features produced by this model  with observations.

We initialize the 2.5D PIC simulations in the solar wind frame of reference with a uniform magnetic field $\mathbf{B}=B_0 \hat{x}$. Both the ion and electron densities are uniform. The initial ion velocity distribution function (VDF) is a single isotropic Maxwellian. The electron VDF is a core-beam isotropic bi-Maxwellian. The core is the solar wind electrons and the beam is the energetic electrons. Their relative drift is along $B_0$: 
\begin{align*}
f_e =\left (\dfrac{m_e}{2\pi k }\right )^{3/2} 
%\end{align*}
%\begin{align*}
 \left [ \dfrac{1-\delta}{T_c^{3/2}}e^{-m_e (v_{e\perp}^2+(v_{ex}-v_{cd}))^2/2k T_c} 
+ \dfrac{\delta}{T_b ^{3/2}} e^{-m_e (v_{e\perp}^2+(v_{ex}-v_{bd}))^2/2k T_b} \right ],
\end{align*}
where $v_{\perp}^2=v_{ey}^2+v_{ez}^2$, $\delta=n_b/n_0$. $n_0$ is the solar wind density and the density normalization unit, $n_b$ is the density of beam electrons. $T_c$ is the temperature of core, $T_b$ is the temperature of beam, $v_{cd}$ is the drift of the core and $v_{bd}$ is the drift of the beam. The drift velocities satisfy $(1-\delta)v_{cd}= -\delta v_{bd}$ to maintain null current. $v_{bd}=12 v_{te} = 60 v_A$, where $v_A$ is the Alfv\'en speed and $v_{te}=\sqrt{k T_c/m_e}$ is the thermal velocity of the core electrons. The energy of these beam electrons will be released and join the core electrons at energy $\sim kT_c$. We choose $\delta=0.1$, because at $\sim 10$ keV, or $\sim 10^3 kT_c$, the superhalo electrons have a density of $\sim 10^{-6}$ of the solar wind density \cite{wang12apjl}, and we assume the kinetic energy flux density of beam $n_b v_{bd}^3/2$ is constant. The speed of light in these simulations is chosen to be $c = 100 v_A$ and the mass ratio is $m_i/m_e=100$. The ion temperature $T_i = T_c$. The boundaries are periodic in both directions with a box size $L_x = L_y =32 d_i$. The total number of cells in each dimension is 10,240 and the total number of particles is $\sim 10^{10}$. The total simulation time $\omega_{pe} t=10,560$. The electric field is normalized by $E_0=v_A B_0/c$. We take $kT_b=2kT_c=0.5 m_i v_A^2$ and $\beta=kT_c/B_0^2=0.25$ estimated from the solar wind $\beta$ observations at 0.3 AU\cite{bou12apj}.

\begin{figure}
\includegraphics[scale=0.7, angle=90, trim=0 0 60 0,clip]{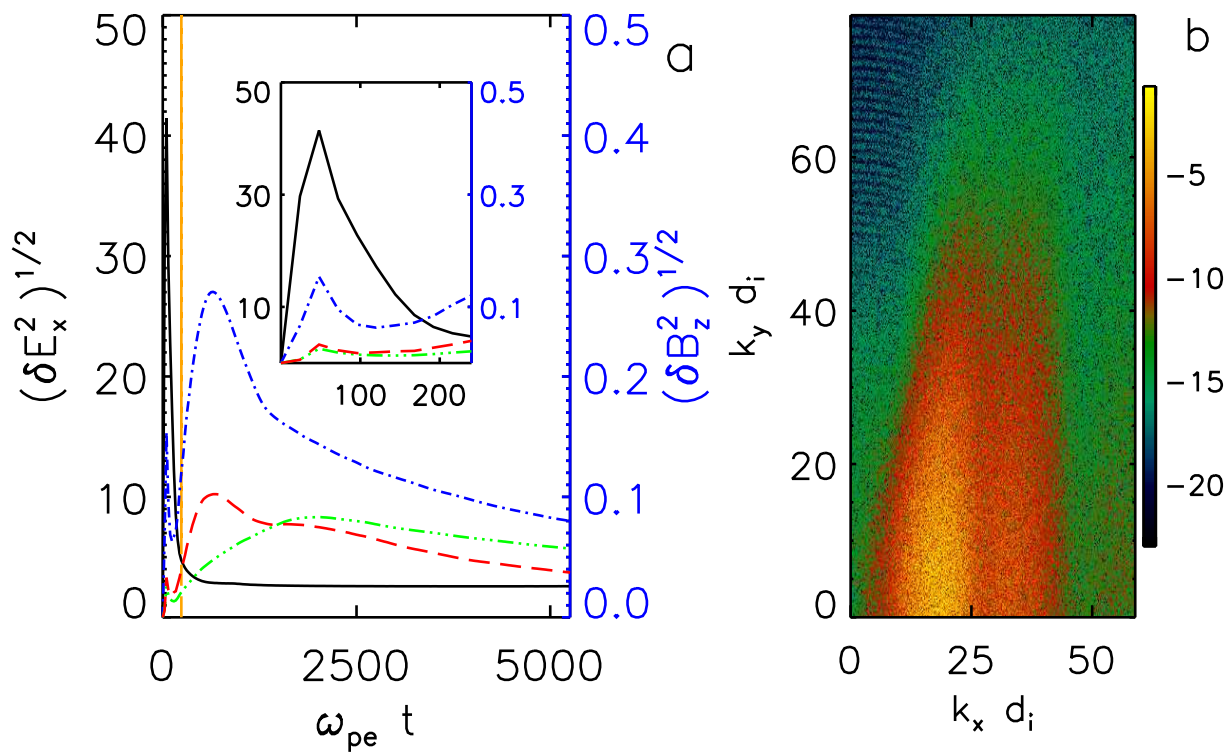} 
\caption{Panel a: Time evolution of energy of $\langle \delta E_x^2\rangle$ (black line) and $\langle \delta B_x^2\rangle$ (red line), $\langle \delta B_y^2\rangle$ (green line), $\langle \delta B_z^2\rangle$ (blue line).  The embedded plot is an expanded view of the time evolution from $\omega_{pe} t=0-230$. The orange line indicates $\omega_{pe} t=230$. Panel b: Power spectrum of $\vert \delta E_x(k_x, k_y)\vert^2$ at $\omega_{pe} t=24$ on a logarithmic scale.}
\label{fluex}
\end{figure}

Electron two-stream instability occurs early at $\omega_{pe}t=24$ as shown in Fig.~\ref{fluex}a, and $\langle \delta E_x^2\rangle$ (solid black line) quickly reaches a peak at $\omega_{pe}t \approx 50$, where $\langle\rangle$ denotes the average over $xy$. At $\omega_{pe} t=200$, the drift of the beams decreases from $60 v_A$ to $\sim 20 v_A$, and $\delta E_x$ decreases by nearly a factor of 20 and then stays nearly constant. The growth rate of the electron two-stream instability at  $\omega_{pe}t=24$ is close to the cold plasma limit of $\gamma_{b}\sim \left (n_b/2n_0\right )^{1/3}\omega_{pe} \sim 0.4 \omega_{pe}$.  The fastest growing mode $k_{f,x} =\omega_{pe}/v_{db}\sim 17/d_i$ is consistent with the spectrum of $\vert \delta E_x(k_x, k_y)\vert^2$
%  on a logarithmic scale 
at  $\omega_{pe}t=24$, as shown in Fig.~\ref{fluex}b.

\begin{figure}
\includegraphics[scale=1,trim=35 0 0 20,clip]{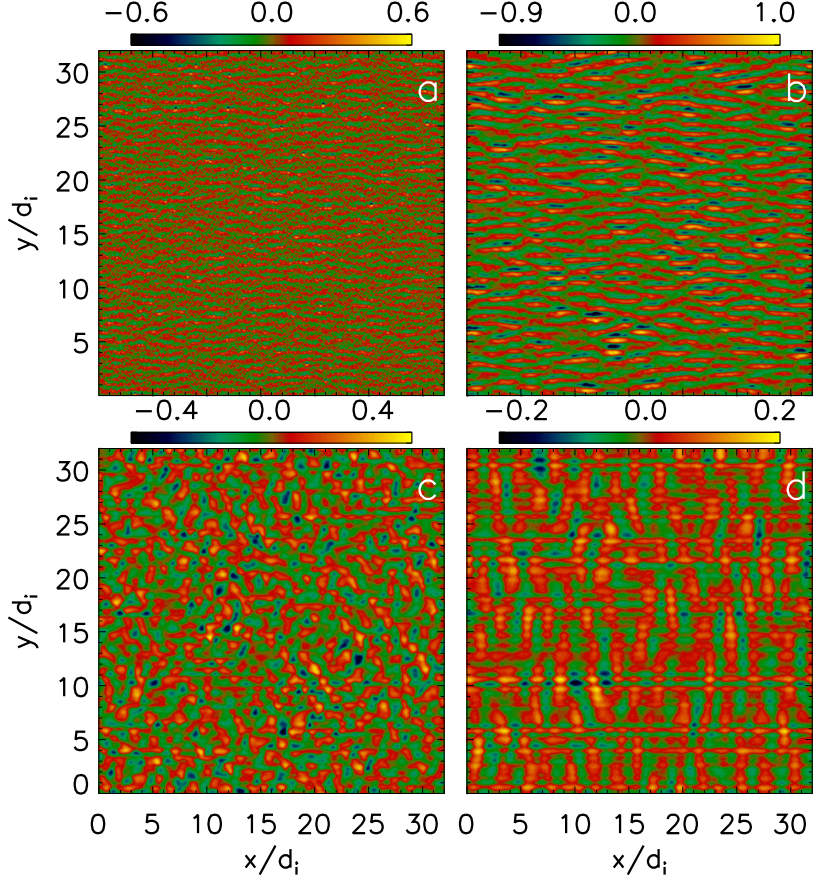} 
\caption{Images of $B_z/B_0$ at $\omega_{pe} t=$ 24 (panel {\bf a}), 480 (panel {\bf b}), 2424 (panel {\bf c}), and 10560 (panel {\bf d}). Please refer to  the movie in the supplement.}
\label{bzall}
\end{figure}

The fast growth of $\delta E_x$ generates an inductive magnetic field $B_z$ that satisfies $ B_z \sim\dfrac{\delta E_x  \Delta y}{c \Delta t}\sim 0.24 B_0$, which is close to the middle value of  $B_z$ shown in Fig.~\ref{bzall}a, where we take $\Delta y \sim \lambda_{f,x}=2\pi/k_{f,x}\sim 3 d_e$, $\Delta t \sim 1/\gamma_b\sim 2.5 \omega_{pe}^{-1}$. 
The middle value of $\delta E_x\sim 20$ during the instability is estimated from Fig.~\ref{fluex}. The internal energy density released per wavelength per is $\sim m_e n_b (\Delta v_{db})^2 \Delta y/(2\omega_{pe} \Delta t)\sim 0.2 n_0 m_i v_A^2 \omega_{pe}^{-1}\lambda_{f,x}^{-1}$ where $\Delta v_{db} \sim 60 v_A$. Around 10\% is converted into magnetic energy $ B^2/8\pi\sim 0.03 n_0 m_i v_A^2 \omega_{pe}^{-1}\lambda_{f,x}^{-1}$ at the end of the two stream instability, while nearly 90\% is converted into the thermal motion of trapped electrons \cite{che13pop}.

\begin{figure}
\includegraphics[scale=1,trim=50 60 30 30,clip]{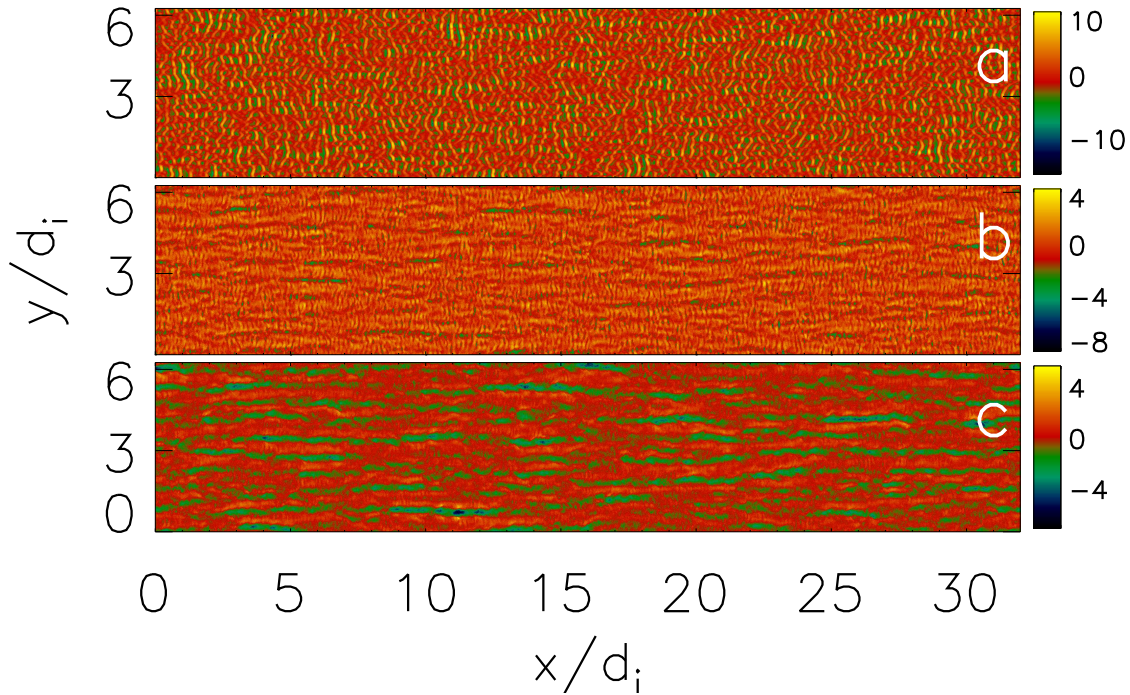} 
\caption{The transition of $j_{ex}$ wave patten when Weibel-like instability occurs. Panel {\bf a}: $j_{ex}$ at $\omega_{pe} t=96$, the late stage of two-stream instability; panel {\bf b}: $j_{ex}$ at $\omega_{pe} t= 168 $, the transition stage from the two-stream instability to the Weibel-like instability; panel {\bf c} at $\omega_{pe} t=240$, the beginning of the Weibel-like instability.}
\label{jex}
\end{figure}

The electric current density  $j_{ex}$ produced by the inductive magnetic field becomes as important as the displacement current when the two-stream instability starts to decay. Then $j_{ex}$ drives a Weibel-like instability that generates nearly non-propagating transverse electromagnetic waves. The variances  $(\delta B_z^2)^{1/2}$ and $(\delta B_x^2)^{1/2}$ in Fig.~\ref{fluex}a reach a second peak at $\omega_{pe} t \approx 672$. The second peak is much higher than the first peak produced by the two-stream instability. The variance $(\delta B_x^2)^{1/2}$ follows $(\delta B_z^2)^{1/2}$ closely, while the variance $(\delta B_y^2)^{1/2}$ reaches its peak at a slightly later time. A significant change from electrostatic waves to transverse electromagnetic waves can be seen in the evolution of $j_{ex}$, shown in Fig.~\ref{jex}. The $j_{ex}$ wave vector induced by the two-stream instability is along $x$. Gradually, the wave vector rotates so that it is parallel to $y$, which indicates the generation of electromagnetic fluctuations in $B_z$ that align along $y$. The wavelength of $B_z$ fluctuations increases to half $d_i$ as seen in Fig.~\ref{bzall}b at $\omega_{pe} t=480$, near the peak of the Weibel-like instability.

The decay of the Weibel-like instability enhances the interactions between the localized currents and the nearly non-propagating transverse electromagnetic waves. This process breaks up the transverse waves and produces randomly propagating waves as shown in Fig~\ref{bzall}c. From $\omega_{pe} t=2400$, the wave-wave interactions dominate the dynamics.  The wave-wave interactions lead to a momentum transfer from perpendicular to parallel magnetic field. As a result, parallel propagating waves appear, which is consistent with the fact that a peak appears in $(\delta B_y^2)^{1/2}$ at $\omega_{pe} t=2400$ (Fig.~\ref{fluex}). Finally at $\omega_{pe} t= 10,560$, nearly perpendicular propagating waves with angle $>89^\circ$ and nearly parallel waves are excited (Fig.~\ref{bzall}d).  

\begin{figure}
\includegraphics[scale=1,trim=50 30 0 20,clip]{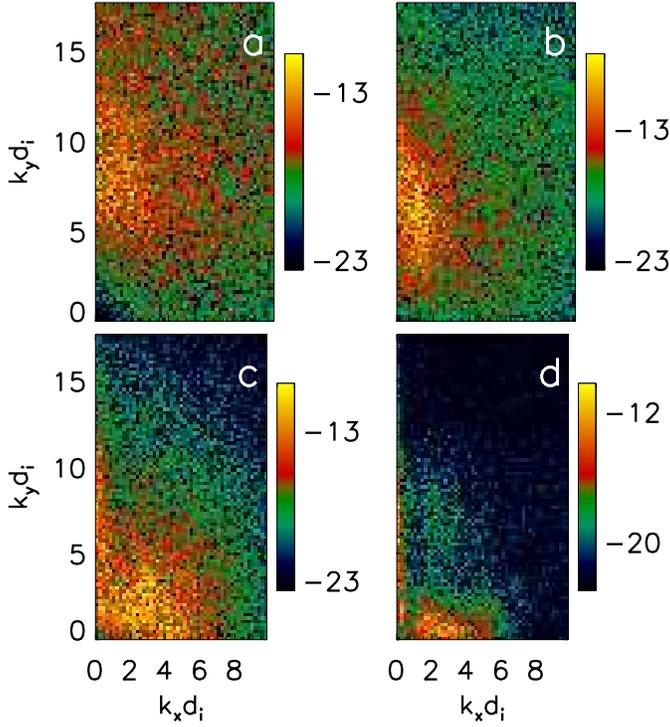} 
\caption{Power spectra $\vert B_z(k_x, k_y)\vert^2$ on logarithmic scale at $\omega_{pe} t=$ 24 (panel a), 480 (panel b), 2424 (panel c), and 10560 (panel d).  }
\label{spec2d}
\end{figure}

The wave-wave interactions drive a bi-directional energy cascade. The perpendicular magnetic wave energy is now transferred from the electron inertial scale back to the ion inertial scale, and the parallel magnetic wave energy is transferred from the ion inertial scale down to the electron inertial scale. The 2D power spectra of $\delta B_z$ at $\omega_{pe} t=$ 24, 480, 2424 and 10,560 are shown in Fig.~\ref{spec2d} (a, b, c, d), respectively. At $\omega_{pe} t=24$, we only see a transverse mode peaked $k_y d_i\sim 10$, i.e., $k_y d_e \sim 1$, which is consistent with the wavelength of the inductive magnetic field $B_z$ that was produced by the two-stream instability.  At $\omega_{pe} t=48$, the Weibel-like instability generates a  transverse electromagnetic magnetic field with longer wavelengths. At $\omega_{pe} t > 2424$, wave-wave interactions occur in which a parallel branch is produced while the wave number of the perpendicular branch decreases.  At the end of the simulation, magnetic powers are concentrated in two branches in the energy spectrum: the nearly perpendicular branch with $k_x d_i<1$, and the parallel branch with $k_y d_i <2$. We study the time evolution of the magnetic components of waves, the results show that both wave types are right-hand polarized. During the evolution, the magnetic wave-wave interactions forms localized thin current sheets with widths from  several $d_e$ to $d_i$. Some of which might be caused by magnetic reconnections (supplementary Fig. 2).

The frequency of the nearly perpendicular wave is around $0.2-0.3 \Omega_i$ where $\Omega_i$ is the ion cyclotron frequency. From the dispersion relation of KAW given by two-fluid equation \cite{sharma2011jgr}
\begin{equation}
\dfrac{\omega^2}{k^2_x v_A^2}=\dfrac{1+k_y^2\rho^2_s}{1+k_y^2 d_e^2}
\end{equation}
\noindent where $\rho_s^2=d_e^2v_{te}^2/v_A^2$, we estimate $k_y d_i < 8$ for $k_x d_i \sim 0.01$ and the electron thermal velocity is larger than the initial velocity $v_{te}^2>T_{c}/m_e=25 v_A^2$. The resulting KAW $k_y d_i$ is consistent with the spectrum shown in Fig.~\ref{spec2d}d. The frequency of the parallel branch is $\omega \sim$ 10 $\Omega_i$ and the wavenumber is $k d_i \sim 3$ at $\omega_{pe} t  = 10,560$ , which satisfies the whistler wave dispersion relation $\omega/\Omega_i=v_A (kd_i)^2 \cos \theta$ \cite{stringer63jne} for $\theta \sim 0$. But at the transition time $\omega_{pe} t = 2424$, $k d_i \sim 4$, then $\theta \sim 45^{0}$ and $k_x d_i\sim 2.5$. Thus the oblique whistler wave evolves to parallel. The ratio of $\delta B_{x}^2/\delta B_{y}^2\sim 1$ at the late stage (Fig.~\ref{fluex}) implies that the turbulent magnetic energy is nearly equally distributed between KAW and whistler wave turbulence.

Three-wave interaction $\mathbf{k}_1\pm \mathbf{k}_2=\mathbf{k}_3$ is the dominant process in wave-wave interactions and leads to the simultaneous generation of KAWs and whistler waves. For perpendicular interactions, the major contribution is from $k_{\perp,1}^{kaw}\pm k_{\perp,2}^{whistler}=k_{\perp,3}^{kaw}$, where $\vert k_{\perp,2}^{whistler}\vert \ll \vert k_{\perp,1}^{kaw}\vert$, thus $k_{\perp}^{kaw}$ moves to smaller wavenumbers and the magnetic energy transfers from small scale to large scale. For parallel interactions, the major contribution is from $k_{\parallel,1}^{kaw}\pm k_{\parallel,2}^{whistler}=k_{\parallel,3}^{whistler}$, where $\vert k_{\parallel,1}^{kaw}\vert \ll \vert k_{\parallel,1}\vert^{whistler}$, thus $k_{\parallel}^{whistler}$ moves to larger wavenumbers and the magnetic energy cascades down to the small scales. 

\begin{figure}
\includegraphics[scale=1,trim=10 20 40 30,clip]{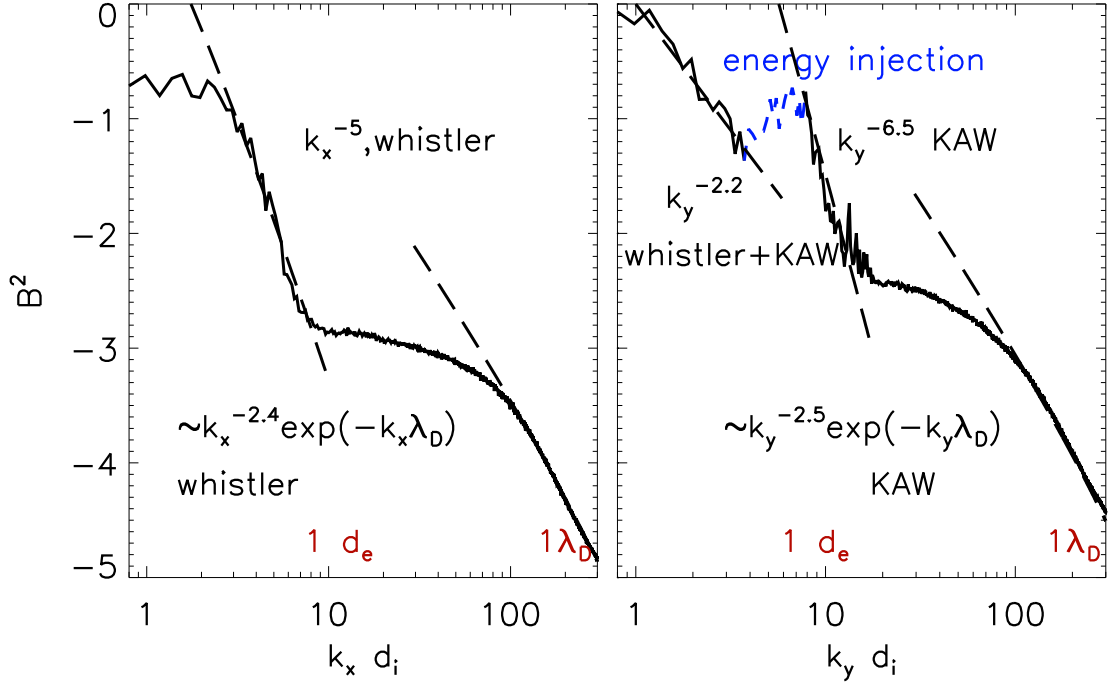} 
\caption{The 1D spectra of $ \delta B^2(k)$ vs. $k_x d_i$  and $k_y d_i$. The blue short-dashed line is the $k_y$ space range for magnetic energy injection. $d_i=2 \rho_i$ and $d_e=2 \rho_e$ in the simulation where $\rho_{i,e}$ are the ion (electron) gyro-radius, thus $k_{x,y} \rho_i =2$ and $k_{x,y} \rho_e =20$}
\label{spec1d}
\end{figure}

In Fig.~\ref{spec1d}, we show 1D power spectra of the magnetic energy $\delta B^2(k)$ vs. $k_x$ (parallel spectrum) and $\delta B^2(k)$ vs. $k_y$ (perpendicular spectrum) at $\omega_{pe} t=10560$. Whistler wave energy cascades from ion to electron scales and it is clear that the contribution to the parallel spectrum in $k_x$ on ion scale is from whistler waves. The perpendicular spectrum has a bump at $k_y d_i \sim 4-7$ corresponding to the relic of magnetic energy injection by the two-stream instability at $\omega_{pe} t=24$ (Fig.~\ref{spec2d} a). Then the wave-wave interactions inversely transfer the KAW energy to ion scale smaller than $d_i$ and generate the whistler waves (Fig.~\ref{spec2d}d). Thus, both the whistler waves and KAWs contribute to the perpendicular spectrum on ion scale while only KAWs contribute to the perpendicular spectrum on electron scale $L_e$ with $d_e> L_e>\rho_e$. The spectrum is much steeper on $L_e$ since the wave-particle interactions are much more stronger. The parallel spectrum on scale smaller than $d_e$  and the perpendicular spectrum on scale smaller than $\rho_e$ suggest exponential decays that imply the dissipation processes are less space and time correlated. The plateaus between the power law and the exponential decays indicate that the energy is accumulated by the strong thermalization. 
 
After $\omega_{pe} t=10560$, the free energy is almost fully released and the induced turbulent scattering produces a nearly isotropic electron halo superposed over the core electrons. The energy exchange between particles and waves reaches balance. The turbulence reaches its new steady state with $P^2 + B^2/8\pi=constant$, $P$ is the total pressure of ions and electrons.The ratio of amplitude of the magnetic fluctuations and background magnetic field is about 0.2 and matches the current observations of solar wind kinetic turbulence.  The decay rate of fluctuations is $\ll 10^{-4} \Omega_{i}$ in the last $3000 \omega_{pe}^{-1}\sim 2\Omega_{i}^{-1}$, estimated from the simulation. This suggests that the kinetic turbulence will be preserved for a long time. If superhalo electrons are produced in the sun, then kinetic turbulence is produced within a few solar radii. The resulting turbulence should be stable enough to travel to 1AU based on the decay rate estimated from our simulation if we take $\Omega_{i} \sim 1$ Hz. 

Our model can naturally explain some important solar wind turbulence observations:  1) Current observations can be compared to the perpendicular power spectrum in Fig.~\ref{spec1d} on scales above $d_e$. The spectral index of -2.2 in our spectrum agrees with those found in observations. In our simulation the KAW as well as the parallel propagating high frequency whistler waves contribute to the power spectrum.  In ref. \cite{alex12apj}, the authors suggest that $\sim 10\%$ of the solar wind data they analysed consists of parallel propagating whistler waves as determined by their right-handed polarization, but more advanced observations are needed. 2) Since the growth rate of two-stream instability is related to $\omega_{pe}$, the spectral breakpoints of KAWs follow the ion inertial length. This agrees with observations of KAW turbulence\cite{Leamon_et_al_2000,perri10apjl,bou12apj}. 3) At the final stage, enhanced by the relic parallel electric field from the two-stream instability, $\langle \vert E_{\parallel}\vert/\vert E_{\perp}\vert\rangle \sim 2-3$, consistent with the observations that the parallel electric field is larger than the perpendicular electric field expected for KAW \cite{mozer13apjl}. 4) During the evolution of the turbulence, microscopic current sheets  with widths varying from several $d_e$ to $d_i$ are produced (supplementary Fig. 2.), consistent with the observations of kinetic scale current sheets discovered in solar wind turbulence\cite{li11prl,perri12prl}. 5) Our simulations show that a nearly isotropic halo is produced at the finale stage. Such halos are observed from 0.3-1 AU in slow wind\cite{pilipp87jgrb} (supplementary Fig.~1). The formation of superhalo requires the electron beam energy extend by more than an order of magnitude higher, rendering computations rather expensive due to the higher $c/v_A$ ratio and higher temporal and spatial resolutions required.

 The observations of kinetic turbulence on electron scale would be more challenging since the power-spectra on electron scale is much steeper and quickly become exponential. The ongoing Magnetospheric Multiscale Mission might be able to detect the kinetic process on electron scale at 1AU. 
 
 It is important to know how other possible turbulent processes in the solar wind affect the KAWs and whistler waves when the they travel to 1 AU.  The index of the power spectra might be affected more while other features produced in our model may be slightly or unaffected: the frequency breakpoints determined by ion inertial length, the enhanced parallel electric field and the electron halo. The current observations of solar wind can reach 0.3 AU. The near future space missions Solar Probe Plus and Solar Orbiter can reach 10 solar radii and hence provide more rigorous constraints on this model.
 
 This model proposed  is motivated by the observations of superhalo. It is encouraging that the model could potentially link the existing observations of solar wind kinetic turbulence, the halo formation, and the electron acceleration and heating processes in solar corona into a coherent picture. More advanced studies will be carried out in the near future.

\begin{acknowledgements}
This research was supported by the NASA Postdoctoral Program at NASA/GSFC administered by Oak Ridge Associated Universities through a contract with NASA. The simulations and analysis were carried out at the NASA Advanced Supercomputing (NAS) facility at the NASA Ames Research Center, and on Kraken at the National Institute for Computation Sciences. 
%Che thanks for the helpful discussions in ``7th Festival de Th\'eorie" held in Aix-en-Provence, France, 2013.
\end{acknowledgements}
\newpage %Just because of unusual number of tables stacked at end
%\bibliography{solarwind}% Produces the bibliography via BibTeX.

\begin{thebibliography}{25}%
\makeatletter
\providecommand \@ifxundefined [1]{%
 \@ifx{#1\undefined}
}%
\providecommand \@ifnum [1]{%
 \ifnum #1\expandafter \@firstoftwo
 \else \expandafter \@secondoftwo
 \fi
}%
\providecommand \@ifx [1]{%
 \ifx #1\expandafter \@firstoftwo
 \else \expandafter \@secondoftwo
 \fi
}%
\providecommand \natexlab [1]{#1}%
\providecommand \enquote  [1]{``#1''}%
\providecommand \bibnamefont  [1]{#1}%
\providecommand \bibfnamefont [1]{#1}%
\providecommand \citenamefont [1]{#1}%
\providecommand \href@noop [0]{\@secondoftwo}%
\providecommand \href [0]{\begingroup \@sanitize@url \@href}%
\providecommand \@href[1]{\@@startlink{#1}\@@href}%
\providecommand \@@href[1]{\endgroup#1\@@endlink}%
\providecommand \@sanitize@url [0]{\catcode `\\12\catcode `\$12\catcode
  `\&12\catcode `\#12\catcode `\^12\catcode `\_12\catcode `\%12\relax}%
\providecommand \@@startlink[1]{}%
\providecommand \@@endlink[0]{}%
\providecommand \url  [0]{\begingroup\@sanitize@url \@url }%
\providecommand \@url [1]{\endgroup\@href {#1}{\urlprefix }}%
\providecommand \urlprefix  [0]{URL }%
\providecommand \Eprint [0]{\href }%
\providecommand \doibase [0]{http://dx.doi.org/}%
\providecommand \selectlanguage [0]{\@gobble}%
\providecommand \bibinfo  [0]{\@secondoftwo}%
\providecommand \bibfield  [0]{\@secondoftwo}%
\providecommand \translation [1]{[#1]}%
\providecommand \BibitemOpen [0]{}%
\providecommand \bibitemStop [0]{}%
\providecommand \bibitemNoStop [0]{.\EOS\space}%
\providecommand \EOS [0]{\spacefactor3000\relax}%
\providecommand \BibitemShut  [1]{\csname bibitem#1\endcsname}%
\let\auto@bib@innerbib\@empty
%</preamble>
\bibitem [{\citenamefont {Leamon}\ \emph {et~al.}(1998)\citenamefont {Leamon},
  \citenamefont {Smith}, \citenamefont {Ness}, \citenamefont {Matthaeus},\ and\
  \citenamefont {Wong}}]{Leamon_et_al_1998a}%
  \BibitemOpen
  \bibfield  {author} {\bibinfo {author} {\bibfnamefont {R.~J.}\ \bibnamefont
  {Leamon}}, \bibinfo {author} {\bibfnamefont {C.~W.}\ \bibnamefont {Smith}},
  \bibinfo {author} {\bibfnamefont {N.~F.}\ \bibnamefont {Ness}}, \bibinfo
  {author} {\bibfnamefont {W.~H.}\ \bibnamefont {Matthaeus}}, \ and\ \bibinfo
  {author} {\bibfnamefont {H.~K.}\ \bibnamefont {Wong}},\ }\href@noop {}
  {\bibfield  {journal} {\bibinfo  {journal} {\jgr}\ }\textbf {\bibinfo
  {volume} {103}},\ \bibinfo {pages} {4775} (\bibinfo {year}
  {1998})}%\BibitemShut {NoStop}%
\bibitem [{\citenamefont {Leamon}\ \emph {et~al.}(1999)\citenamefont {Leamon},
  \citenamefont {Smith}, \citenamefont {Ness},\ and\ \citenamefont
  {Wong}}]{Leamon_et_al_1999}%
  \BibitemOpen
  \bibfield  {author} {\bibinfo {author} {\bibfnamefont {R.~J.}\ \bibnamefont
  {Leamon}}, \bibinfo {author} {\bibfnamefont {C.~W.}\ \bibnamefont {Smith}},
  \bibinfo {author} {\bibfnamefont {N.~F.}\ \bibnamefont {Ness}}, \ and\
  \bibinfo {author} {\bibfnamefont {H.~K.}\ \bibnamefont {Wong}},\ }\href@noop
  {} {\bibfield  {journal} {\bibinfo  {journal} {\jgr}\ }\textbf {\bibinfo
  {volume} {104}},\ \bibinfo {pages} {22,331} (\bibinfo {year}
  {1999})}%\BibitemShut {NoStop}%
\bibitem [{\citenamefont {Leamon}\ \emph {et~al.}(2000)\citenamefont {Leamon},
  \citenamefont {Matthaeus}, \citenamefont {Smith}, \citenamefont {Zank},
  \citenamefont {Mullan},\ and\ \citenamefont {Oughton}}]{Leamon_et_al_2000}%
  \BibitemOpen
  \bibfield  {author} {\bibinfo {author} {\bibfnamefont {R.~J.}\ \bibnamefont
  {Leamon}}, \bibinfo {author} {\bibfnamefont {W.~H.}\ \bibnamefont
  {Matthaeus}}, \bibinfo {author} {\bibfnamefont {C.~W.}\ \bibnamefont
  {Smith}}, \bibinfo {author} {\bibfnamefont {G.~P.}\ \bibnamefont {Zank}},
  \bibinfo {author} {\bibfnamefont {D.~J.}\ \bibnamefont {Mullan}}, \ and\
  \bibinfo {author} {\bibfnamefont {S.}~\bibnamefont {Oughton}},\ }\href@noop
  {} {\bibfield  {journal} {\bibinfo  {journal} {\apj}\ }\textbf {\bibinfo
  {volume} {537}},\ \bibinfo {pages} {1054} (\bibinfo {year}
  {2000})}%\BibitemShut {NoStop}%
\bibitem [{\citenamefont {{Bale}}\ \emph {et~al.}(2005)\citenamefont {{Bale}},
  \citenamefont {{Kellogg}}, \citenamefont {{Mozer}}, \citenamefont
  {{Horbury}},\ and\ \citenamefont {{Reme}}}]{bale05prl}%
  \BibitemOpen
  \bibfield  {author} {\bibinfo {author} {\bibfnamefont {S.~D.}\ \bibnamefont
  {{Bale}}}, \bibinfo {author} {\bibfnamefont {P.~J.}\ \bibnamefont
  {{Kellogg}}}, \bibinfo {author} {\bibfnamefont {F.~S.}\ \bibnamefont
  {{Mozer}}}, \bibinfo {author} {\bibfnamefont {T.~S.}\ \bibnamefont
  {{Horbury}}}, \ and\ \bibinfo {author} {\bibfnamefont {H.}~\bibnamefont
  {{Reme}}},\ }\href {\doibase 10.1103/PhysRevLett.94.215002} {\bibfield
  {journal} {\bibinfo  {journal} {\prl}\ }\textbf {\bibinfo {volume} {94}},\
  \bibinfo {eid} {215002} (\bibinfo {year} {2005})},\ \Eprint
  {http://arxiv.org/abs/arXiv:physics/0503103} {arXiv:physics/0503103}
  %\BibitemShut {NoStop}%
\bibitem [{\citenamefont {{Sahraoui}}\ \emph {et~al.}(2009)\citenamefont
  {{Sahraoui}}, \citenamefont {{Goldstein}}, \citenamefont {{Robert}},\ and\
  \citenamefont {{Khotyaintsev}}}]{sah09prl}%
  \BibitemOpen
  \bibfield  {author} {\bibinfo {author} {\bibfnamefont {F.}~\bibnamefont
  {{Sahraoui}}}, \bibinfo {author} {\bibfnamefont {M.~L.}\ \bibnamefont
  {{Goldstein}}}, \bibinfo {author} {\bibfnamefont {P.}~\bibnamefont
  {{Robert}}}, \ and\ \bibinfo {author} {\bibfnamefont {Y.~V.}\ \bibnamefont
  {{Khotyaintsev}}},\ }\href {\doibase 10.1103/PhysRevLett.102.231102}
  {\bibfield  {journal} {\bibinfo  {journal} {\prl}\ }\textbf {\bibinfo
  {volume} {102}},\ \bibinfo {eid} {231102} (\bibinfo {year}
  {2009})}%\BibitemShut {NoStop}%
\bibitem [{\citenamefont {{Kiyani}}\ \emph {et~al.}(2009)\citenamefont
  {{Kiyani}}, \citenamefont {{Chapman}}, \citenamefont {{Khotyaintsev}},
  \citenamefont {{Dunlop}},\ and\ \citenamefont {{Sahraoui}}}]{kiyani09prl}%
  \BibitemOpen
  \bibfield  {author} {\bibinfo {author} {\bibfnamefont {K.~H.}\ \bibnamefont
  {{Kiyani}}}, \bibinfo {author} {\bibfnamefont {S.~C.}\ \bibnamefont
  {{Chapman}}}, \bibinfo {author} {\bibfnamefont {Y.~V.}\ \bibnamefont
  {{Khotyaintsev}}}, \bibinfo {author} {\bibfnamefont {M.~W.}\ \bibnamefont
  {{Dunlop}}}, \ and\ \bibinfo {author} {\bibfnamefont {F.}~\bibnamefont
  {{Sahraoui}}},\ }\href {\doibase 10.1103/PhysRevLett.103.075006} {\bibfield
  {journal} {\bibinfo  {journal} {\prl}\ }\textbf {\bibinfo {volume} {103}},\
  \bibinfo {eid} {075006} (\bibinfo {year} {2009})},\ \Eprint
  {http://arxiv.org/abs/0906.2830} {arXiv:0906.2830 [physics.plasm-ph]}
  %\BibitemShut {NoStop}%
\bibitem [{\citenamefont {{Alexandrova}}\ \emph {et~al.}(2009)\citenamefont
  {{Alexandrova}}, \citenamefont {{Saur}}, \citenamefont {{Lacombe}},
  \citenamefont {{Mangeney}}, \citenamefont {{Mitchell}}, \citenamefont
  {{Schwartz}},\ and\ \citenamefont {{Robert}}}]{alex09prl}%
  \BibitemOpen
  \bibfield  {author} {\bibinfo {author} {\bibfnamefont {O.}~\bibnamefont
  {{Alexandrova}}}, \bibinfo {author} {\bibfnamefont {J.}~\bibnamefont
  {{Saur}}}, \bibinfo {author} {\bibfnamefont {C.}~\bibnamefont {{Lacombe}}},
  \bibinfo {author} {\bibfnamefont {A.}~\bibnamefont {{Mangeney}}}, \bibinfo
  {author} {\bibfnamefont {J.}~\bibnamefont {{Mitchell}}}, \bibinfo {author}
  {\bibfnamefont {S.~J.}\ \bibnamefont {{Schwartz}}}, \ and\ \bibinfo {author}
  {\bibfnamefont {P.}~\bibnamefont {{Robert}}},\ }\href {\doibase
  10.1103/PhysRevLett.103.165003} {\bibfield  {journal} {\bibinfo  {journal}
  {\prl}\ }\textbf {\bibinfo {volume} {103}},\ \bibinfo {eid} {165003}
  (\bibinfo {year} {2009})},\ \Eprint {http://arxiv.org/abs/0906.3236}
  {arXiv:0906.3236 [physics.plasm-ph]} %\BibitemShut {NoStop}%
\bibitem [{\citenamefont {{Salem}}\ \emph {et~al.}(2012)\citenamefont
  {{Salem}}, \citenamefont {{Howes}}, \citenamefont {{Sundkvist}},
  \citenamefont {{Bale}}, \citenamefont {{Chaston}}, \citenamefont {{Chen}},\
  and\ \citenamefont {{Mozer}}}]{salem12apjl}%
  \BibitemOpen
  \bibfield  {author} {\bibinfo {author} {\bibfnamefont {C.~S.}\ \bibnamefont
  {{Salem}}}, \bibinfo {author} {\bibfnamefont {G.~G.}\ \bibnamefont
  {{Howes}}}, \bibinfo {author} {\bibfnamefont {D.}~\bibnamefont
  {{Sundkvist}}}, \bibinfo {author} {\bibfnamefont {S.~D.}\ \bibnamefont
  {{Bale}}}, \bibinfo {author} {\bibfnamefont {C.~C.}\ \bibnamefont
  {{Chaston}}}, \bibinfo {author} {\bibfnamefont {C.~H.~K.}\ \bibnamefont
  {{Chen}}}, \ and\ \bibinfo {author} {\bibfnamefont {F.~S.}\ \bibnamefont
  {{Mozer}}},\ }\href {\doibase 10.1088/2041-8205/745/1/L9} {\bibfield
  {journal} {\bibinfo  {journal} {\apjl}\ }\textbf {\bibinfo {volume} {745}},\
  \bibinfo {eid} {L9} (\bibinfo {year} {2012})}%\BibitemShut {NoStop}%
\bibitem [{\citenamefont {{Podesta}}(2013)}]{pod13solphys}%
  \BibitemOpen
  \bibfield  {author} {\bibinfo {author} {\bibfnamefont {J.~J.}\ \bibnamefont
  {{Podesta}}},\ }\href {\doibase 10.1007/s11207-013-0258-z} {\bibfield
  {journal} {\bibinfo  {journal} {\solphys}\ } (\bibinfo {year} {2013}),\
  10.1007/s11207-013-0258-z}%\BibitemShut {NoStop}%
\bibitem [{\citenamefont {{Perri}}\ \emph {et~al.}(2010)\citenamefont
  {{Perri}}, \citenamefont {{Carbone}},\ and\ \citenamefont
  {{Veltri}}}]{perri10apjl}%
  \BibitemOpen
  \bibfield  {author} {\bibinfo {author} {\bibfnamefont {S.}~\bibnamefont
  {{Perri}}}, \bibinfo {author} {\bibfnamefont {V.}~\bibnamefont {{Carbone}}},
  \ and\ \bibinfo {author} {\bibfnamefont {P.}~\bibnamefont {{Veltri}}},\
  }\href {\doibase 10.1088/2041-8205/725/1/L52} {\bibfield  {journal} {\bibinfo
   {journal} {\apjl}\ }\textbf {\bibinfo {volume} {725}},\ \bibinfo {pages}
  {L52} (\bibinfo {year} {2010})}%\BibitemShut {NoStop}%
\bibitem [{\citenamefont {{Bourouaine}}\ \emph {et~al.}(2012)\citenamefont
  {{Bourouaine}}, \citenamefont {{Alexandrova}}, \citenamefont {{Marsch}},\
  and\ \citenamefont {{Maksimovic}}}]{bou12apj}%
  \BibitemOpen
  \bibfield  {author} {\bibinfo {author} {\bibfnamefont {S.}~\bibnamefont
  {{Bourouaine}}}, \bibinfo {author} {\bibfnamefont {O.}~\bibnamefont
  {{Alexandrova}}}, \bibinfo {author} {\bibfnamefont {E.}~\bibnamefont
  {{Marsch}}}, \ and\ \bibinfo {author} {\bibfnamefont {M.}~\bibnamefont
  {{Maksimovic}}},\ }\href {\doibase 10.1088/0004-637X/749/2/102} {\bibfield
  {journal} {\bibinfo  {journal} {\apj}\ }\textbf {\bibinfo {volume} {749}},\
  \bibinfo {eid} {102} (\bibinfo {year} {2012})}%\BibitemShut {NoStop}%
\bibitem [{\citenamefont {{Howes}}\ \emph {et~al.}(2008)\citenamefont
  {{Howes}}, \citenamefont {{Dorland}}, \citenamefont {{Cowley}}, \citenamefont
  {{Hammett}}, \citenamefont {{Quataert}}, \citenamefont {{Schekochihin}},\
  and\ \citenamefont {{Tatsuno}}}]{howes08prl}%
  \BibitemOpen
  \bibfield  {author} {\bibinfo {author} {\bibfnamefont {G.~G.}\ \bibnamefont
  {{Howes}}}, \bibinfo {author} {\bibfnamefont {W.}~\bibnamefont {{Dorland}}},
  \bibinfo {author} {\bibfnamefont {S.~C.}\ \bibnamefont {{Cowley}}}, \bibinfo
  {author} {\bibfnamefont {G.~W.}\ \bibnamefont {{Hammett}}}, \bibinfo {author}
  {\bibfnamefont {E.}~\bibnamefont {{Quataert}}}, \bibinfo {author}
  {\bibfnamefont {A.~A.}\ \bibnamefont {{Schekochihin}}}, \ and\ \bibinfo
  {author} {\bibfnamefont {T.}~\bibnamefont {{Tatsuno}}},\ }\href {\doibase
  10.1103/PhysRevLett.100.065004} {\bibfield  {journal} {\bibinfo  {journal}
  {\prl}\ }\textbf {\bibinfo {volume} {100}},\ \bibinfo {eid} {065004}
  (\bibinfo {year} {2008})},\ \Eprint {http://arxiv.org/abs/0711.4355}
  {arXiv:0711.4355} %\BibitemShut {NoStop}%
\bibitem [{\citenamefont {{TenBarge}}\ \emph {et~al.}(2013)\citenamefont
  {{TenBarge}}, \citenamefont {{Howes}},\ and\ \citenamefont
  {{Dorland}}}]{ten13apj}%
  \BibitemOpen
  \bibfield  {author} {\bibinfo {author} {\bibfnamefont {J.~M.}\ \bibnamefont
  {{TenBarge}}}, \bibinfo {author} {\bibfnamefont {G.~G.}\ \bibnamefont
  {{Howes}}}, \ and\ \bibinfo {author} {\bibfnamefont {W.}~\bibnamefont
  {{Dorland}}},\ }\href {\doibase 10.1088/0004-637X/774/2/139} {\bibfield
  {journal} {\bibinfo  {journal} {\apj}\ }\textbf {\bibinfo {volume} {774}},\
  \bibinfo {eid} {139} (\bibinfo {year} {2013})}%\BibitemShut {NoStop}%
\bibitem [{\citenamefont {{Lin}}(1997)}]{lin97conf}%
  \BibitemOpen
  \bibfield  {author} {\bibinfo {author} {\bibfnamefont {R.~P.}\ \bibnamefont
  {{Lin}}},\ }in\ \href@noop {} {\emph {\bibinfo {booktitle} {Scientific Basis
  for Robotic Exploration Close to the Sun}}},\ \bibinfo {series} {AIP
  Conference Proceedings}, Vol.\ \bibinfo {volume} {385},\ \bibinfo
  {organization} {AIP}\ (\bibinfo  {publisher} {American Institute of
  Physics},\ \bibinfo {year} {1997})\ pp.\ \bibinfo {pages}
  {25--32}%\BibitemShut {NoStop}%
\bibitem [{\citenamefont {{Wang}}\ \emph {et~al.}(2012)\citenamefont {{Wang}},
  \citenamefont {{Lin}}, \citenamefont {{Salem}}, \citenamefont {{Pulupa}},
  \citenamefont {{Larson}}, \citenamefont {{Yoon}},\ and\ \citenamefont
  {{Luhmann}}}]{wang12apjl}%
  \BibitemOpen
  \bibfield  {author} {\bibinfo {author} {\bibfnamefont {L.}~\bibnamefont
  {{Wang}}}, \bibinfo {author} {\bibfnamefont {R.~P.}\ \bibnamefont {{Lin}}},
  \bibinfo {author} {\bibfnamefont {C.}~\bibnamefont {{Salem}}}, \bibinfo
  {author} {\bibfnamefont {M.}~\bibnamefont {{Pulupa}}}, \bibinfo {author}
  {\bibfnamefont {D.~E.}\ \bibnamefont {{Larson}}}, \bibinfo {author}
  {\bibfnamefont {P.~H.}\ \bibnamefont {{Yoon}}}, \ and\ \bibinfo {author}
  {\bibfnamefont {J.~G.}\ \bibnamefont {{Luhmann}}},\ }\href {\doibase
  10.1088/2041-8205/753/1/L23} {\bibfield  {journal} {\bibinfo  {journal}
  {\apjl}\ }\textbf {\bibinfo {volume} {753}},\ \bibinfo {eid} {L23} (\bibinfo
  {year} {2012})}%\BibitemShut {NoStop}%
\bibitem [{\citenamefont {{Parker}}(1988)}]{parker88apj}%
  \BibitemOpen
  \bibfield  {author} {\bibinfo {author} {\bibfnamefont {E.~N.}\ \bibnamefont
  {{Parker}}},\ }\href {\doibase 10.1086/166485} {\bibfield  {journal}
  {\bibinfo  {journal} {\apj}\ }\textbf {\bibinfo {volume} {330}},\ \bibinfo
  {pages} {474} (\bibinfo {year} {1988})}%\BibitemShut {NoStop}%
\bibitem [{\citenamefont {{Gary}}(1993)}]{gary93book}%
  \BibitemOpen
  \bibfield  {author} {\bibinfo {author} {\bibfnamefont {S.~P.}\ \bibnamefont
  {{Gary}}},\ }\href@noop {} {\emph {\bibinfo {title} {Theory of Space Plasma
  Microinstabilities, by S.~Peter Gary, pp.~193.~ISBN 0521431670.~Cambridge,
  UK: Cambridge University Press, September 1993.}}}\ (\bibinfo {year}
  {1993})%\BibitemShut {NoStop}%
\bibitem [{\citenamefont {{Pilipp}}\ \emph {et~al.}(1987)\citenamefont
  {{Pilipp}}, \citenamefont {{Muehlhaeuser}}, \citenamefont {{Miggenrieder}},
  \citenamefont {{Rosenbauer}},\ and\ \citenamefont
  {{Schwenn}}}]{pilipp87jgrb}%
  \BibitemOpen
  \bibfield  {author} {\bibinfo {author} {\bibfnamefont {W.~G.}\ \bibnamefont
  {{Pilipp}}}, \bibinfo {author} {\bibfnamefont {K.-H.}\ \bibnamefont
  {{Muehlhaeuser}}}, \bibinfo {author} {\bibfnamefont {H.}~\bibnamefont
  {{Miggenrieder}}}, \bibinfo {author} {\bibfnamefont {H.}~\bibnamefont
  {{Rosenbauer}}}, \ and\ \bibinfo {author} {\bibfnamefont {R.}~\bibnamefont
  {{Schwenn}}},\ }\href {\doibase 10.1029/JA092iA02p01103} {\bibfield
  {journal} {\bibinfo  {journal} {\jgr}\ }\textbf {\bibinfo {volume} {92}},\
  \bibinfo {pages} {1103} (\bibinfo {year} {1987})}%\BibitemShut {NoStop}%
\bibitem [{\citenamefont {Che}\ \emph {et~al.}(2013)\citenamefont {Che},
  \citenamefont {Drake}, \citenamefont {Swisdak},\ and\ \citenamefont
  {Goldstein}}]{che13pop}%
  \BibitemOpen
  \bibfield  {author} {\bibinfo {author} {\bibfnamefont {H.}~\bibnamefont
  {Che}}, \bibinfo {author} {\bibfnamefont {J.~F.}\ \bibnamefont {Drake}},
  \bibinfo {author} {\bibfnamefont {M.}~\bibnamefont {Swisdak}}, \ and\
  \bibinfo {author} {\bibfnamefont {M.~L.}\ \bibnamefont {Goldstein}},\ }\href
  {\doibase 10.1063/1.4811137} {\bibfield  {journal} {\bibinfo  {journal}
  {\pop}\ }\textbf {\bibinfo {volume} {20}},\ \bibinfo {eid} {061205} (\bibinfo
  {year} {2013})}%\BibitemShut {NoStop}%
\bibitem [{\citenamefont {Sharma}\ and\ \citenamefont
  {Kumar}(2011)}]{sharma2011jgr}%
  \BibitemOpen
  \bibfield  {author} {\bibinfo {author} {\bibfnamefont {R.~P.}\ \bibnamefont
  {Sharma}}\ and\ \bibinfo {author} {\bibfnamefont {S.}~\bibnamefont {Kumar}},\
  }\href@noop {} {\bibfield  {journal} {\bibinfo  {journal} {\jgr}\ }\textbf
  {\bibinfo {volume} {116}},\ \bibinfo {pages} {A03103} (\bibinfo {year}
  {2011})}%\BibitemShut {NoStop}%
\bibitem [{\citenamefont {{Stringer}}(1963)}]{stringer63jne}%
  \BibitemOpen
  \bibfield  {author} {\bibinfo {author} {\bibfnamefont {T.~E.}\ \bibnamefont
  {{Stringer}}},\ }\href {\doibase 10.1088/0368-3281/5/2/304} {\bibfield
  {journal} {\bibinfo  {journal} {Journal of Nuclear Energy}\ }\textbf
  {\bibinfo {volume} {5}},\ \bibinfo {pages} {89} (\bibinfo {year}
  {1963})}%\BibitemShut {NoStop}%
\bibitem [{\citenamefont {{Alexandrova}}\ \emph {et~al.}(2012)\citenamefont
  {{Alexandrova}}, \citenamefont {{Lacombe}}, \citenamefont {{Mangeney}},
  \citenamefont {{Grappin}},\ and\ \citenamefont {{Maksimovic}}}]{alex12apj}%
  \BibitemOpen
  \bibfield  {author} {\bibinfo {author} {\bibfnamefont {O.}~\bibnamefont
  {{Alexandrova}}}, \bibinfo {author} {\bibfnamefont {C.}~\bibnamefont
  {{Lacombe}}}, \bibinfo {author} {\bibfnamefont {A.}~\bibnamefont
  {{Mangeney}}}, \bibinfo {author} {\bibfnamefont {R.}~\bibnamefont
  {{Grappin}}}, \ and\ \bibinfo {author} {\bibfnamefont {M.}~\bibnamefont
  {{Maksimovic}}},\ }\href {\doibase 10.1088/0004-637X/760/2/121} {\bibfield
  {journal} {\bibinfo  {journal} {\apj}\ }\textbf {\bibinfo {volume} {760}},\
  \bibinfo {eid} {121} (\bibinfo {year} {2012})},\ \Eprint
  {http://arxiv.org/abs/1212.0412} {arXiv:1212.0412 [astro-ph.SR]} %\BibitemShut
  {NoStop}%
\bibitem [{\citenamefont {{Mozer}}\ and\ \citenamefont
  {{Chen}}(2013)}]{mozer13apjl}%
  \BibitemOpen
  \bibfield  {author} {\bibinfo {author} {\bibfnamefont {F.~S.}\ \bibnamefont
  {{Mozer}}}\ and\ \bibinfo {author} {\bibfnamefont {C.~H.~K.}\ \bibnamefont
  {{Chen}}},\ }\href {\doibase 10.1088/2041-8205/768/1/L10} {\bibfield
  {journal} {\bibinfo  {journal} {\apjl}\ }\textbf {\bibinfo {volume} {768}},\
  \bibinfo {eid} {L10} (\bibinfo {year} {2013})},\ \Eprint
  {http://arxiv.org/abs/1304.1189} {arXiv:1304.1189 [physics.space-ph]}
  %\BibitemShut {NoStop}%
\bibitem [{\citenamefont {{Li}}\ \emph {et~al.}(2011)\citenamefont {{Li}},
  \citenamefont {{Miao}}, \citenamefont {{Hu}},\ and\ \citenamefont
  {{Qin}}}]{li11prl}%
  \BibitemOpen
  \bibfield  {author} {\bibinfo {author} {\bibfnamefont {G.}~\bibnamefont
  {{Li}}}, \bibinfo {author} {\bibfnamefont {B.}~\bibnamefont {{Miao}}},
  \bibinfo {author} {\bibfnamefont {Q.}~\bibnamefont {{Hu}}}, \ and\ \bibinfo
  {author} {\bibfnamefont {G.}~\bibnamefont {{Qin}}},\ }\href {\doibase
  10.1103/PhysRevLett.106.125001} {\bibfield  {journal} {\bibinfo  {journal}
  {\prl}\ }\textbf {\bibinfo {volume} {106}},\ \bibinfo {eid} {125001}
  (\bibinfo {year} {2011})}%\BibitemShut {NoStop}%
\bibitem [{\citenamefont {{Perri}}\ \emph {et~al.}(2012)\citenamefont
  {{Perri}}, \citenamefont {{Goldstein}}, \citenamefont {{Dorelli}},\ and\
  \citenamefont {{Sahraoui}}}]{perri12prl}%
  \BibitemOpen
  \bibfield  {author} {\bibinfo {author} {\bibfnamefont {S.}~\bibnamefont
  {{Perri}}}, \bibinfo {author} {\bibfnamefont {M.~L.}\ \bibnamefont
  {{Goldstein}}}, \bibinfo {author} {\bibfnamefont {J.~C.}\ \bibnamefont
  {{Dorelli}}}, \ and\ \bibinfo {author} {\bibfnamefont {F.}~\bibnamefont
  {{Sahraoui}}},\ }\href {\doibase 10.1103/PhysRevLett.109.191101} {\bibfield
  {journal} {\bibinfo  {journal} {\prl}\ }\textbf {\bibinfo {volume} {109}},\
  \bibinfo {eid} {191101} (\bibinfo {year} {2012})}%\BibitemShut {NoStop}%
\end{thebibliography}
%

\end{document}